\documentclass[manuscript,screen]{acmart} 

\AtBeginDocument{%
  }

\setcopyright{acmlicensed}
\copyrightyear{2026}
\acmYear{2026}
\acmDOI{XXXXXXX.XXXXXXX}
\acmJournal{CSUR}

\begin{document}


\title{Preference-Based Learning in Audio Applications: A Systematic Analysis}

\author{Aaron Broukhim}
\email{aabroukh@ucsd.edu}
\orcid{0009-0007-4092-8048}
\affiliation{%
  \institution{UC San Diego}
  \city{La Jolla}
  \state{CA}
  \country{USA}
}

\author{Yiran Shen}
\email{jes038@ucsd.edu}
\orcid{0009-0006-1663-9740}
\affiliation{%
  \institution{UC San Diego}
  \city{La Jolla}
  \state{CA}
  \country{USA}
}

\author{Prithviraj Ammanabrolu}
\email{prithvi@ucsd.edu}
\orcid{0000-0003-1950-0454} 
\affiliation{%
  \institution{UC San Diego}
  \city{La Jolla}
  \state{CA}
  \country{USA}
}

\author{Nadir Weibel}
\email{weibel@ucsd.edu}
\orcid{0000-0002-3457-4227}
\affiliation{%
  \institution{UC San Diego}
  \city{La Jolla}
  \state{CA}
  \country{USA}
}

\renewcommand{\shortauthors}{Broukhim et al.}

\begin{abstract}
    Despite the parallel challenges that audio and text domains face in evaluating generative model outputs, preference learning remains remarkably underexplored in audio applications. Through a PRISMA-guided systematic review of approximately 500 papers, we find that only 30 (6\%) apply preference learning to audio tasks. Our analysis reveals a field in transition: pre-2021 works focused on emotion recognition using traditional ranking methods (rankSVM), while post-2021 studies have pivoted toward generation tasks employing modern RLHF frameworks. We identify three critical patterns: (1) the emergence of multi-dimensional evaluation strategies combining synthetic, automated, and human preferences; (2) inconsistent alignment between traditional metrics (WER, PESQ) and human judgments across different contexts; and (3) convergence on multi-stage training pipelines that combine reward signals. Our findings suggest that while preference learning shows promise for audio, particularly in capturing subjective qualities like naturalness and musicality, the field requires standardized benchmarks, higher-quality datasets, and systematic investigation of how temporal factors unique to audio impact preference learning frameworks.
\end{abstract}



\received{17 November 2025}

\maketitle

\section{Introduction}
With the proliferation of generative AI models, attempts to properly evaluate their outputs proved difficult. Early examples in text models found evaluation metrics, such as BLEU \cite{papineni2002bleu} or ROUGE \cite{lin2004rouge}---which measure n-gram overlap with reference texts---to often misalign with human evaluations of their systems. Applied to machine translation, summarization, and image caption generation, these metrics typically compare generated texts to a reference gold standard to determine the quality of texts. For instance, BLEU rewards exact word matches while penalizing valid paraphrases \cite{reiter2018structured}, ROUGE can favor repetitive text \cite{schluter2017limits}, and even eventual LLM-as-Judge frameworks fail to properly evaluate creativity \cite{chakrabarty2024art}. Works like InstructGPT thus turned to preference learning to develop a means to evaluate these more qualitative areas like helpfulness \cite{ouyang2022training}. 

Preference learning converts the difficult and low agreement task of qualitative ratings of generative model outputs to a ranking task. Given two or more outputs for the same query, an annotator can rank the outputs with respect to some criteria outlined by researchers (e.g. helpfulness). However, while humans struggle to consistently rate quality on absolute scales, we can reliably judge which of two outputs is better \cite{carterette2008here, wester2015we}. This has the dual benefit of making data collection easier and expands the range of signals for model fine-tuning. In audio, this approach is particularly valuable: preference learning allows systems to capture human judgments about sound quality, style, and naturalness through paired comparisons—such as which sample sounds more pleasant or expressive—rather than relying solely on numerical scores.

Metrics often used to evaluate audio signals suffer similar fates to text metrics. PESQ \cite{rix2001perceptual}, WER \cite{morris2004and}, MuLan embeddings \cite{huang2022mulan} and others don't necessarily align with human evaluations. PESQ fails for modern neural vocoders \cite{manocha2020differentiable}, WER can improve while naturalness degrades in speech synthesis \cite{anastassiou2024seed}, and MuLan explicitly isn't aligned with human evaluations of music generation \cite{cideron2024musicrl}. Despite this parallel, very little work applies preference learning frameworks to the audio domain. 

Our systematic review of approximately 500 papers reveals that only 30 (6\%) apply preference learning to audio, suggesting a nascent field. This paper provides the first comprehensive synthesis of preference learning in audio through a PRISMA-guided systematic review. We analyze 30 works spanning from 2010 to 2025, bridging early ranking approaches with modern Reinforcement Learning from Human Feedback (RLHF) methods, and identify critical patterns in how preference learning adapts to audio's unique challenges, from temporal perception to emotion recognition to musical creativity. After introducing the foundations of preference learning and audio evaluation---including key distinctions that guide effective data-collection design---(Section~\ref{sect:background}), we describe our systematic methodology (Section~\ref{sect:methods}), then synthesize findings across tasks, data approaches, and evaluation methods (Section~\ref{sect:results}), before discussing notable findings and future directions for this emerging field (Section~\ref{sect:discussion}).

To the best of our knowledge, there is no prior systematic review of preference based learning methods in audio. Existing preference learning surveys strictly remain in the text modality \cite{chaudhari2025rlhf, kaufmann2024survey}. Audio specific surveys like those on deep learning \cite{zhu2021deep} or deep reinforcement learning for audio \cite{latif2101survey} do not consider preference learning in their frameworks. Our work bridges this gap by providing the first systematic analysis that spans both pre-modern ranking approaches (like rankSVM) and modern RLHF frameworks specifically for audio applications.

\section{Background}
\label{sect:background}

\subsection{Preference Learning Fundamentals}
\begin{figure}[t!]
    \centering    \includegraphics[width=\textwidth]{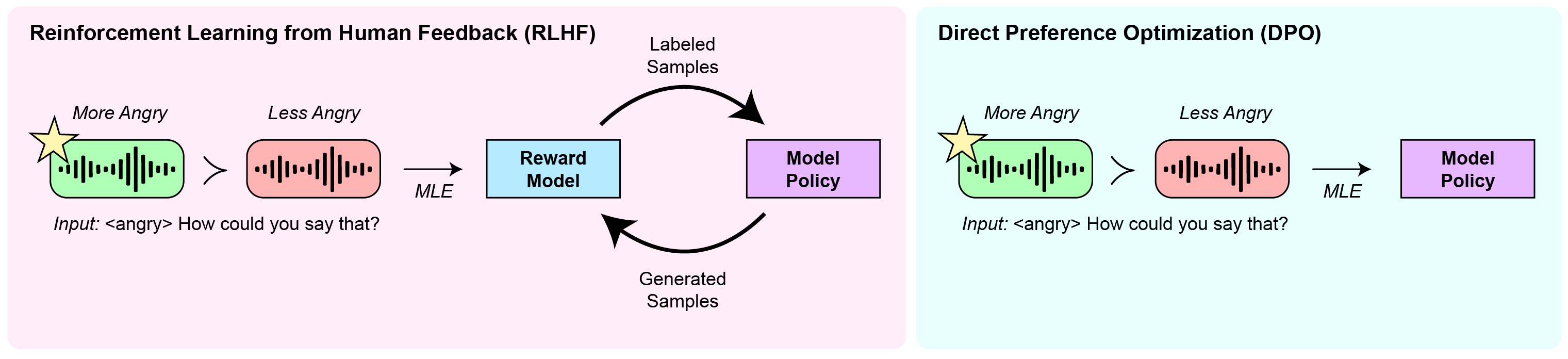}
    \vspace{-2em}
    \caption{A visual representation of RLHF and DPO training loops. Both pipelines implement RLHF; the key difference is whether the reward model is explicit (left) or implicit (right). MLE stand for Maximum Likelihood Estimation.}
    \Description{Comparison of RLHF and DPO training pipelines. RLHF (left) uses an explicit reward model trained on labeled preference samples to guide policy training. DPO (right) directly optimizes the policy using preference pairs without a separate reward model. Both use voice sample preferences (angry vs. less angry) as input.}
    \vspace{-1em}
\label{fig:rlhf-dpo_visual}
\end{figure}

Preference learning builds on the idea that models can be aligned to human judgments by learning from pairwise comparisons between outputs. In modern systems, this is achieved through frameworks such as Reinforcement Learning from Human Feedback (RLHF) and Direct Preference Optimization (DPO), which differ primarily in whether they rely on an explicit reward model or directly optimize a policy from preference data. Figure~\ref{fig:rlhf-dpo_visual} provides an overview of these training pipelines.

Stiennon et al. \cite{stiennon2020learning} present one of the first applications of the RLHF framework. For the task of summarization, their model improves significantly with respect to coverage, coherence, accuracy, and overall performance relative to their non-fine-tuned baseline. Notably their improved performance on this subjective evaluation did \textit{not} align with an improvement in ROUGE score. This motivated follow-up works on larger scale and cross-task applications within the text domain.

A desirable attribute of this paradigm is the simplicity and high annotation agreement of ranking data. Multiple works have shown this higher agreement rate across tasks and domains \cite{carterette2008here, wester2015we}. This is often attributed to the way humans conduct non-referenced based annotations, in which they'll rate relative to previously annotated instances instead of towards some absolute "true" score. Some works do convert non-reference annotations to ranked pairs or lists, but these tend to be weaker signals than direct comparisons \cite{metallinou2013annotation}. Understanding the distinction between reference-based and non-reference annotations is crucial because the source of preference data directly determines annotation agreement, reward-signal quality, and ultimately the effectiveness of preference-learning systems.

We distinguish between reference-based comparisons (where instances are evaluated relative to each other) and non-reference annotations (where instances are labeled independently). Reference-based comparisons include sequential annotations, direct A/B comparisons, and absolute score estimation when multiple instances are presented at once. Sequential annotations are those in which an annotator labels changes over time to some audio clip, the reference is considered to be the audio segment prior to the current annotation. A/B comparisons, or those with more than 2 comparisons like list rankings, inherently have a reference at time of annotation. These can be binary A>B tasks or absolute scores. Absolute scores are only considered reference-based when presented with some additional reference data point. Non-reference annotations are those in which a label is applied to an instance independently of other instances.  Once collected, rankings can be leveraged to either train a reward model or directly fine-tune an existing base model. 

Reward models are typically trained using the Bradley-Terry model with pairwise comparison data \cite{bradley1952rank,christiano2017deep} with additional approaches for list-wise ranking tasks (more than 2) \cite{plackett1975analysis,luce1959individual, cao2007learning}. Reward models aren't typically trained from scratch, but start off as some relevant base model that is fine-tuned using the Bradley-Terry model. The Bradley-Terry model outputs the probability a given instance $i$ is preferred to -- or better than -- $j$, depending on the application. This is outlined via the equation below: 
\begin{equation}
P(i \succ j) = \sigma(r_i - r_j) = \frac{1}{1 + \exp(-(r_i - r_j))}
\end{equation}
where $r_i$ and $r_j$ represent the reward or utility scores for instances $i$ and $j$ respectively, and $i \succ j$ denotes that $i$ is preferred over $j$. During training, the model learns to assign scalar rewards such that the Bradley-Terry probabilities match the observed preferences in the training data, with the goal of generalizing these learned preferences to unseen instances.

Once a reward model is trained, its outputs can be leveraged to fine-tune some base model. This can be tangibly represented as follows: we collect n pairs of data from humans saying text x was better than text y given the same prompt, we use said n pairs to train a reward model, and finally the reward model is used to fine-tune a base model, guiding it to generate outputs that align with the preferences learned from the x and y comparisons.

Formally, the typical training pipeline is as follows. First, some model is trained from scratch on copious amounts of domain-specific data (unsupervised), this model is then further fine-tuned on more structured task-specific data (supervised), and lastly an RL-type training loop is used as outlined above to align the model to human preferences. Given a reward model, various algorithms are used to fine-tune the supervised base model (REINFORCE \cite{zhang2021sample}, PPO \cite{schulman2017proximal, ouyang2022training}, etc.), but we leave those specifics to other works \cite{casper2023open}. More recently, Direct Preference Optimization (DPO) \cite{rafailov2023direct} bypasses the need for an explicit reward model by directly optimizing the policy using preference data through a modified loss function. 

While preference learning has its use cases, there are still challenges associated with it as a learning framework. Dataset collection can be expensive as large amounts of annotations are needed \cite{ouyang2022training}, training can be unstable \cite{ziegler2019fine} although recent works are attempting to stabilize the training process \cite{shao2024deepseekmath}, and models are prone to reward hacking \cite{gao2023scaling}. Of these challenges, reward hacking is particularly problematic \cite{pan2022effects}. Reward hacking occurs when a model exploits flaws in the reward function to achieve high rewards without actually improving on the intended task. Attempts to mitigate this vary, with KL-divergence constraints being the most prominent. KL-constraints prevent a model from deviating from its supervised fine-tuned version and thus generating outputs within a more reasonable distribution, preventing extreme deviations from basic coherency. More formally, KL-divergence acts as a regularization term that penalizes deviations between the learned policy and a reference policy, preventing over-optimization to the reward signal.

\subsection{Why Audio is Suited for Preference Learning}
Tasks in the audio domain are often multi-faceted in their evaluation needs: speech synthesis must output the correct semantic content while expressing that information with appropriate intonation and prosody;  music generation needs to do more than simply predict the most common next note---it requires holistic qualitative evaluation of how 'good' the output sounds. Similarly, text-to-audio generation involves temporal relationships (e.g., a cat meowing while a car drives by) that traditional metrics fail to capture \cite{liao2024baton}.

Traditional audio evaluation relies on metrics that each capture specific aspects of quality. For semantic accuracy, Word Error Rate (WER) and Character Error Rate (CER) measure transcription precision at word (e.g. in English) and character (e.g. in Mandarin) levels respectively, counting insertions, deletions, and substitutions. While designed for speech recognition, these metrics are often repurposed for synthesis evaluation by transcribing generated speech and comparing it to input text.

Acoustic similarity metrics attempt to quantify how closely generated audio matches reference recordings. Speaker Similarity (SIM) uses cosine similarity between speaker embeddings to ensure voice consistency in cloning and adaptation tasks. Frechet Audio Distance (FAD) \cite{kilgour2018fr} takes a broader approach, measuring the statistical distance between distributions of audio embeddings from real and generated samples.

Moving toward perceptual alignment, metrics like Perceptual Evaluation of Speech Quality (PESQ) \cite{rix2001perceptual} and Non-Intrusive Objective Speech Quality Assessment (NISQA) \cite{mittag2021nisqa} aim to predict human quality judgments. PESQ models the human auditory system to compare degraded and reference signals, while NISQA operates without references, making it practical for real-world applications. CLAP scores bridge acoustic and semantic evaluation by assessing how well audio corresponds to text descriptions using multimodal embeddings.

We note the similarities of these metrics to early text metrics like BLEU and ROUGE, often simply calculating a distance measure to some reference gold standard signal. As discussed earlier, in generative models these metrics, while initially useful, reach a limitation when qualitative evaluations are needed like naturalness, musicality, or confidence. To address these gaps in objective evaluation, the field still relies heavily on human evaluation.
 
Among human evaluation methods, Multiple Stimuli with Hidden Reference and Anchor (MUSHRA) \cite{itu20031534} provides a controlled framework for subjective assessment, having annotators compare multiple systems simultaneously against hidden references and anchors to ensure reliable discrimination between systems. Other generative audio tasks typically employ task-specific mean opinion scores (MOS) to evaluate their end systems.

These diverse metrics often provide conflicting signals: a system might achieve low WER while sounding robotic, or high PESQ scores while failing to capture emotional expression. This multidimensional evaluation challenge, combined with the inherent subjectivity of audio perception, makes preference learning particularly suitable for audio applications. 

\vspace{1em}
The variation in annotation formats, data-collection strategies, and reward-modeling choices described above—together with challenges such as reward hacking and instability—makes it difficult to compare existing preference-learning approaches for audio. To address this lack of cohesion and provide clarity on emerging patterns, we conduct a systematic review of existing work. In the next section we detail the methodology behind our survey.

\section{Methods}
\label{sect:methods}
\begin{figure}[t]
    \centering
    \includegraphics[width=1.0\textwidth]{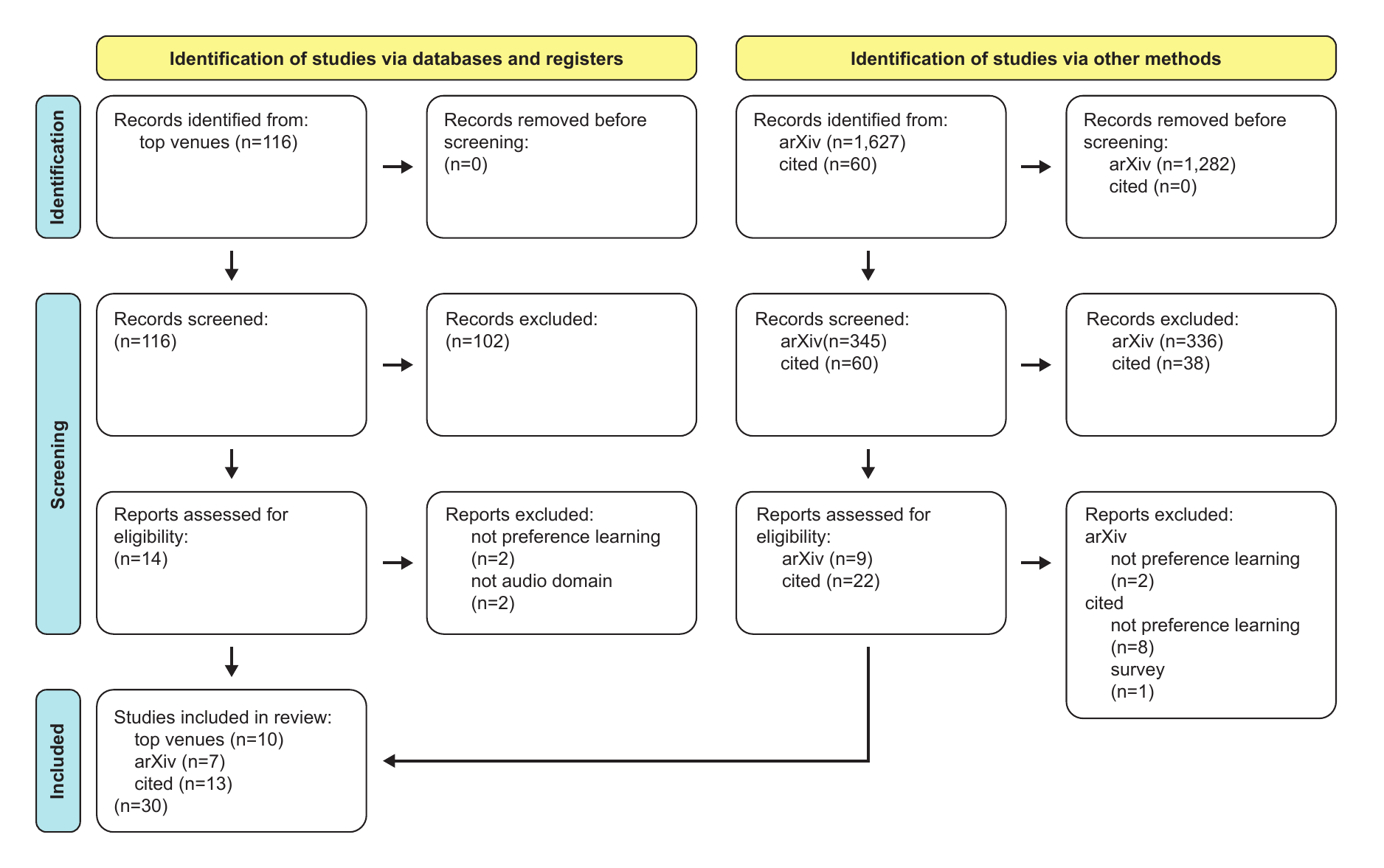}
    \caption{PRISMA-inspired diagram illustrating the literature screening process. Papers were excluded for one of three reasons: (1) absence of a preference learning framework, (2) lack of focus on the audio domain, or (3) being a survey.}
    \label{fig:PRISMA}
    \Description{PRISMA systematic review flow diagram tracking study selection across four stages. Left pathway shows identification from academic databases (116 records from top venues, 10 studies included). Right pathway shows identification from arXiv and citation searching (1,687 records identified, 20 studies included after screening and exclusions for non-preference learning and non-audio domain studies). Total of 30 studies included in final review.}
\end{figure}

To construct a focused and comprehensive survey of preference learning in the audio domain, we employed a multi-stage literature retrieval and screening process inspired by the PRISMA framework \cite{page2021prisma}. This involved a keyword-based search across top-tier machine learning and audio research venues, as well as arXiv for non-peer-reviewed but timely work. Papers were filtered by publication date, keyword relevance, venue prominence, and citation-based thresholds. Each candidate paper was assessed against predefined inclusion criteria centered on audio-based preference learning. This process yielded 30 works for inclusion, comprising published papers, arXiv preprints, and works cited by the selected literature. The full selection workflow is summarized in Figure \ref{fig:PRISMA}.

\subsection{Keyword Selection}
To identify works relevant to preference learning in the audio domain, we conducted a keyword-based search. We started with a broad initial keyword set: ("Preference" OR "Feedback" OR "Ranking") AND ("Audio" OR "Conversation" OR "Music" OR "Speech" OR "Sound"), which returned a substantial number of irrelevant papers. We then iteratively refined our search criteria through multiple rounds of pruning and expansion.

We removed keywords that appeared too general and validated this decision by reviewing abstracts of papers that would be excluded. Conversely, while reading relevant papers, we identified candidate keywords for inclusion and validated these by reviewing abstracts of papers that would be newly retrieved. When removal or addition of a keyword changed the result set by more than 100 papers, we reviewed a random sample of those papers to assess whether the keyword should be retained. This iterative process continued until successive refinement rounds yielded no new candidate keywords.

The final search string was: ("audio" OR "music" OR "acoustic") AND ("preference learning" OR "RLHF" OR "DPO" OR "KTO"), applied to the full text of papers. Additional keywords were tested but excluded due to low relevance or high false positive rates, including: "verbal," "spoken," "speech," "ordinal," "alignment," "utterance," "ASR" (automatic speech recognition), and "LALM" (large audio language models).

Given the rapid pace of research in preference learning and audio machine learning, we limited our search to works published from 2020 onward, the year RLHF began seeing modern applications \cite{stiennon2020learning}. Literature retrieval was conducted on March 19, 2025 using Publish or Perish \cite{harzing2007pop} with Google Scholar as the source.

\subsection{Published Works}
As preference learning falls within the field of machine learning, we focused on the top three machine learning venues ranked by h5-index, according to Google Scholar \cite{google_scholar_hindex}. These venues are: Neural Information Processing Systems (NeurIPS; h5-index = 337), the International Conference on Learning Representations (ICLR; h5-index = 304), and the International Conference on Machine Learning (ICML; h5-index = 268).

To ensure coverage of domain-specific research, we also included the top venues in the acoustics and sound category, again based on h5-index values from Google Scholar \cite{google_scholar_hindex}. These venues are: the IEEE International Conference on Acoustics, Speech and Signal Processing (ICASSP; h5-index = 129), the Conference of the International Speech Communication Association (INTERSPEECH; h5-index = 111), and the IEEE/ACM Transactions on Audio, Speech, and Language Processing (IEEE/ACM TASLP; h5-index = 74).

\subsection{Non-Published Works}
Beyond the six peer-reviewed venues, we applied the same keyword search to arXiv. Although arXiv is not peer-reviewed, it has become a de facto outlet for timely or industry-oriented research \cite{ginsparg2021lessons, sutton2017popularity}, particularly in fast-evolving areas. Including arXiv enables us to capture emerging and pre-publication works that are often critical to ongoing developments.

Due to the volume of arXiv publications containing the selected keywords since 2020, we included only those within the top 20th percentile by citation count (e.g., “$\ge$ 32 cites for 2022 papers”). This threshold was determined based on the observation that 9\% (10/116) of papers from the venue-based search were potentially relevant, and we assumed an 11\% false-negative rate. Exact citation thresholds by year are provided in the Appendix, Table \ref{tab:top20-screening}.

\subsection{Cited Works}
During the screening and review process, if a cited paper was potentially relevant to this review, it was noted for potential inclusion. This occurred during the screening and the eventual full read-through detailed below. Using this process, we find 60 potentially relevant cited works. We define a cited paper as potentially relevant if its title or description within the text that cites it seems like it may follow the criteria outlined below.

\subsection{Final Selection and Inclusion}
In total, this search yielded 116 published papers, 345 arXiv papers, and 60 cited works for potential inclusion in the literature review. For each paper, the abstract, introduction, and conclusion were read to assess eligibility. A paper was deemed eligible if it satisfied the following criteria: 

\begin{enumerate}
    \item it pertains to preference learning
    \item it is applied within the audio domain\\
    \emph{and}
    \item it is not a survey or review article
\end{enumerate}
For a full breakdown of non-inclusion reasoning see Table \ref{tab:screening_reasons} in the Appendix.

We consider a paper to be related to preference learning if it involves the ranking or A/B comparison of two or more audio clips. Numeric ratings (e.g., 1–5 MOS) were included only when the rating was converted to rankings (e.g. 5>3 therefore A>B). These preferences may be derived either from human judgments or generated synthetically.  We further include works that may not have comparisons but a RL training loop for an audio model. Studies that used preferences solely for evaluation, with no learning component, were excluded. Multi-modal works including audio were classified as within the audio domain. All eligibility decisions were made according to this protocol by a single reviewer.

Considering the above criteria, we narrowed down: published papers from 116 to 14, arXiv papers from 345 to 9, and cited works from 60 to 22. During the review process no duplicates were identified. These identified papers were then read in full for final considerations of inclusion. During the full read-through, we further exclude 4 published works, 2 arXiv works, and 9 cited works. 30 total works were thus included in this survey, 10 from the venue search, 7 from arXiv, and 13 cited works. 

Figure \ref{fig:PRISMA} summarizes the literature screening and selection process following the PRISMA framework. We provide the BibTex for all reviewed and included works at the following \href{https://drive.google.com/drive/folders/1UwyWoSpem-aFixIQLBd6tgELl5R-4Q6I?usp=sharing}{link\footnote{After publication these bibtex files can be made accessible as supplemental material}}.

\section{Results}
\label{sect:results}
We present an analysis of 30 selected works spanning 15 years of preference learning in audio applications (see Table~\ref{tab:conf_dist}). We identify work from 2010 to 2020 as pre-modern preference learning, which is characterized by ranking based supervision, rankSVM style models, and early neural architectures for emotion recognition. We identify work from 2021 onward as modern preference learning, which is defined by reinforcement learning from preference signals and is primarily applied to generative audio tasks. Our analysis reveals three key findings: 

\begin{enumerate}
    \item  A dramatic shift from classification-focused emotion recognition to generation tasks, with audio generation studies emerging exclusively after 2021
    
    \item The convergence on multi-dimensional evaluation strategies
    
    \item Context-dependent relationships between traditional metrics and human judgments, where metrics like WER show variable correlation across different applications, highlighting the need for multi-metric evaluation frameworks 
\end{enumerate}

\begin{table}[h]
\centering
  \caption{Distribution of Selected Works by Venue, indicating representative years}
  \label{tab:conf_dist}
  \small
  \begin{tabular}{@{}l c l l@{}}
    \toprule
    \textbf{Venue} & \textbf{Count} & \textbf{Ref.} & \textbf{Representative Years}\\
    \midrule
    IEEE/ACM TASLP & 2 & \cite{yang2010ranking, parthasarathy2016using} & 2010 -- 2016 \\
    INTERSPEECH & 8 & \cite{cao2012combining, lotfian2016retrieving, parthasarathy2018preference, dong2020pyramid, jayawardena2020ordinal, liu2021reinforcement, naini2023preference, wu2023interval} & 2012 -- 2023 \\
    Comput. Speech Lang. & 1 & \cite{cao2015speaker} & 2015 \\
    ICASSP & 7 & \cite{lotfian2016practical, parthasarathy2017ranking, han2020ordinal, naini2023unsupervised, kumar2025using, nagpal2025speech, gao2025emo} & 2016 -- 2025 \\
    IEEE TAC & 2 & \cite{lopes2017modelling, lei2023audio} & 2017 -- 2023 \\
    ICML & 3 & \cite{chumbalov2020scalable, cideron2024musicrl, wu2025adaptive} & 2020 -- 2025 \\
    IUI & 1 & \cite{zhou2021interactive} & 2021 \\
    IJCAI & 1 & \cite{liao2024baton} & 2024 \\
    NeurIPS & 1 & \cite{zhang2024speechalign} & 2024 \\
    arXiv & 4 & \cite{anastassiou2024seed, chu2024qwen2, huang2025step, luo2025openomni} & 2024 -- 2025 \\
    \bottomrule
  \end{tabular}
\end{table}

\subsection{Tasks}
To better understand the utility of the analyzed work, we categorized them by primary task, across the publication timeline. Early works \cite{cao2012combining, lotfian2016practical, parthasarathy2016using} focused on emotion recognition tasks, primarily in speech, with notable exceptions including music \cite{yang2010ranking} and soundscape emotion recognition \cite{lopes2017modelling}. More recent studies extend emotion recognition to multiple modalities \cite{lei2023audio}. Table \ref{tab:task_distribution} details the categorization of the primary tasks across the 30 selected works. Of the 30 included works $\sim$3\% (1) address audio retrieval, $\sim$17\% (5) address speech and voice processing, 30\% (9) address audio generation, and 50\% (15) address emotion analysis. For a detailed breakdown of specific emotions considered in both recognition and generation tasks, please see Table \ref{tab:emotions_considered_distribution} in the Appendix.

\begin{table}[t]
\centering
  \caption{Task Distribution in Identified Works}
  \label{tab:task_distribution}
  \small
    \begin{tabular}{@{}l c l l} 
    \toprule
    \textbf{Task Category} & \textbf{Studies} & \textbf{Representative Works} & \textbf{Representative Years}\\
    \midrule
    \textbf{Emotion Analysis} & \textbf{15} & & \textbf{2010 -- 2023} \\
    \quad Speech Emotion Recognition & 12 & \cite{cao2012combining, cao2015speaker, lotfian2016practical, lotfian2016retrieving, parthasarathy2016using, parthasarathy2018preference, parthasarathy2017ranking, han2020ordinal, jayawardena2020ordinal, naini2023unsupervised, naini2023preference, wu2023interval} & 2012 -- 2023\\
    \quad Music Emotion Recognition & 1 &  \cite{yang2010ranking} & 2010\\
    \quad Soundscape Emotion Recognition & 1 & \cite{lopes2017modelling} & 2017 \\
    \quad Multi-Modal Emotion Recognition & 1 & \cite{lei2023audio} & 2023 \\
    \addlinespace
    
    \textbf{Audio Generation} & \textbf{9} & & \textbf{2021 -- 2025} \\
    \quad Speech Emotion Generation & 4 & \cite{liu2021reinforcement, anastassiou2024seed, luo2025openomni, gao2025emo} & 2021 -- 2025 \\
    \quad Music Generation & 3 & \cite{zhou2021interactive, cideron2024musicrl, wu2025adaptive} & 2021 -- 2025 \\
    \quad Text-to-Audio Generation & 1 & \cite{liao2024baton} & 2024 \\
    \addlinespace

    \textbf{Speech \& Voice Processing} & \textbf{5} & & \textbf{2020 -- 2025} \\
    \quad Speech Quality Assessment & 2 & \cite{dong2020pyramid, kumar2025using} & 2020 -- 2025 \\
    \quad Speech Model Alignment & 3 & \cite{chu2024qwen2, zhang2024speechalign, huang2025step} & 2024 -- 2025 \\
    \quad Automatic Speech Recognition & 1 & \cite{nagpal2025speech} & 2025 \\
    \addlinespace

    \textbf{Audio Retrieval} & \textbf{1} & & \textbf{2020} \\
    \quad Comparison Based Search & 1 & \cite{chumbalov2020scalable} & 2020 \\
    \bottomrule
  \end{tabular}
\end{table}

Audio generation has shown the fastest growth, with 9 of the analyzed studies in the time span of just 5 years. We note the increase of publications in machine learning specific venues specifically in those 5 years, as seen in Table \ref{tab:conf_dist}. These applications span expressive emotion speech generation \cite{anastassiou2024seed, gao2025emo}, music generation \cite{cideron2024musicrl}, and text-to-audio generation \cite{liao2024baton}. Recently, preference learning frameworks traditionally used for model alignment have been applied to speech models \cite{zhang2024speechalign, chu2024qwen2, huang2025step}. Additional niche applications from the 2020-2025 include speech quality assessment \cite{dong2020pyramid}, audio retrieval \cite{chumbalov2020scalable}, and automatic speech recognition \cite{nagpal2025speech}.

\subsection{Pre-Modern Preference Learning (Learning to Rank)}
Early works investigated emotion recognition, predominantly applied to speech as a ranking task instead of rating task. These works consistently found that ranking data improved models' abilities to recognize emotions \cite{yang2010ranking, cao2012combining, cao2015speaker} although ensemble methods including both ranking and rating-based classifiers also performed well \cite{cao2012combining}. Ranking also improved cross-corpus SER, showing gains in generalization relative to rating predictions \cite{naini2023unsupervised}. These analyses, mostly conducted from 2010 to 2016, were applied to support vector machines and referred to as rankSVMs. The rankSVM architecture was eventually found to be less effective at speech emotion recognition tasks than RankNet \cite{parthasarathy2017ranking} before various deep learning architectures became more prominent around 2020. 

In addition to novel domain applications such as horror soundscaping with the same rankSVM learning framework \cite{lopes2017modelling}, deep learning architectures became more prominent \cite{huang2025step, kumar2025using, liao2024baton} with applications such as speech quality assessment \cite{dong2020pyramid} and audio generation \cite{liu2021reinforcement}. The first speech emotion recognition paper to use a deep learning framework identified was published in 2020 and converted the k-class classification problem to an aggregate 'k-1'-binary ranking problem applied to the convolutional neural network, VGGish \cite{han2020ordinal}. Speech quality assessment was also attempted by predicting subjective, mean opinion scores (MOS) of speech audio quality \cite{dong2020pyramid}. This type of MOS prediction would eventually be considered a reward signal to finetune generative models in more modern frameworks \cite{kumar2025using}.

\subsection{Modern Preference Learning}
\begin{table}[h]
\centering
  \caption{Modern Audio Preference Learning Paradigms}
  \label{tab:preference_learning_distribution}
  \small
    \begin{tabular}{@{}l l l l l} 
    \toprule
    \textbf{Work} & \textbf{Base Model} & \textbf{Reward Signal} & \textbf{Pref. Type} & \textbf{Training Framework}\\
    \midrule
    \cite{liu2021reinforcement} & Encoder-Decoder & SER & NA & REINFORCE \\
    \addlinespace

    \cite{anastassiou2024seed} & Transformer & SIM, WER, SER & NA & REINFORCE \\
    \addlinespace

    \cite{nagpal2025speech} & Transformer & \parbox[t]{5cm}{WER \\ LLM Meaning Pres. Score} & NA & PPO \\
    \addlinespace

    \cite{wu2025adaptive} & Transformer & \parbox[t]{5cm}{Cos. Sim. (Melody-Chord Embedding)\\ Discriminator (Real vs Random Pair)} & Synthetic & REINFORCE \\
    \addlinespace

    \cite{gao2025emo} & Transformer & Gen. Target Emotion vs Gen. Non-Target & Synthetic & DPO \\
    \addlinespace

    \cite{luo2025openomni} & Transformer & Gen. Target Emotion vs Gen. Non-Target  & Synthetic & DPO \\
    \addlinespace

    \cite{zhang2024speechalign} & Transformer & Real Speech Tokenizer vs Synth. Speech Tokenizer & Synthetic & \parbox[t]{3cm}{DPO \\ PPO} \\
    \addlinespace

    \cite{cideron2024musicrl} & Transformer & \parbox[t]{5cm}{Cos. Sim. (Audio-Text Embedding)\\ Acoustic Quality MOS\\ Subjective Binary Music Preferences} & Synth.+Real & REINFORCE \\
    \addlinespace

    \cite{huang2025step} & Transformer & \parbox[t]{5cm}{LLM-as-Judge\\ Human Ratings (1-5) of Instruction, Naturalness, \& Safety} & Synth.+Real & PPO \\
    \addlinespace

    \cite{kumar2025using} & LSTM & Audio Quality MOS & Real & PPO \\
    \addlinespace
    

    \cite{liao2024baton} & Diffusion & Binary Preferences & Real & Reward MSE \\
    \addlinespace
    
    \bottomrule
  \end{tabular}

  \footnotesize{Note: NA means the signal is not available as it is an automated metric, not a preference label.}
\end{table}

Modern preference-learning frameworks, like RLHF and DPO, have historically been used for generative model instruction alignment in text \cite{ouyang2022training}. Audio models are no exception, with some papers using preference learning to align \textit{what} is said \cite{chu2024qwen2, cideron2024musicrl}. We note the base model architectures and training paradigms in Table \ref{tab:preference_learning_distribution}. The pre-modern works above can be viewed as potential reward models for a feedback training loop. Unique to audio generation is the use of preference learning for \textit{how} audio sounds instead of only semantic content. 

The evolution of preference learning in audio generation reveals several convergent patterns that transcend individual implementations. The universal adoption of KL constraints across different algorithms (REINFORCE, PPO, DPO) suggests fundamental challenges in maintaining generation quality while optimizing for preferences. The persistent combination of multiple reward signals (whether classification-based, synthetic, or human) indicates that no single evaluation paradigm captures the full complexity of audio quality.

Among RLHF algorithms, REINFORCE, PPO, and DPO have been implemented for speech tasks. While DPO is notably easier to implement than the alternatives, all three have proven effective for speech emotion generation \cite{anastassiou2024seed}. SpeechAlign \cite{zhang2024speechalign} provides a direct comparison: PPO achieved the highest MOS improvement after a single iteration, while DPO showed progressive gains through iterations 2 and 3 before declining at iteration 4. The computational efficiency of DPO enabled multiple iterations, whereas PPO was limited to one. Notably, the performance gains from iterations 1→2 and 2→3 exceeded those from the initial iteration (0→1). Additionally, while SpeechAlign \cite{zhang2024speechalign} found that while both DPO and PPO outperformed supervised fine-tuning, PPO demonstrated superior generalization to unseen datasets. The application of these techniques is also broadening, with BATON \cite{liao2024baton}  pioneering the use of a reward-modulated loss to apply preference learning directly to a diffusion model.

\subsubsection{Preference Reward Models}
A common strategy for creating a reward signal is to measure the semantic similarity between embeddings. For instance, to improve text adherence in music generation, Cideron et al. \cite{cideron2024musicrl} calculated the cosine similarity between text and audio embeddings. Similarly, ReaLchords \cite{wu2025adaptive} improved musical harmony by using the cosine similarity of melody and chord embeddings as a reward. This work also trained a separate discriminator model to reward synchronization by distinguishing between real and randomly paired melodies and chords.

Another prominent approach involves training models directly on human judgments. Cideron et al. \cite{cideron2024musicrl} trained a user preference model on pairwise rankings from human annotators, while Step-Audio \cite{huang2025step} used an LLM-as-judge framework to generate preference data for fine-tuning a reward model. Other work uses scalar feedback; for example, Cideron et al. \cite{cideron2024musicrl} and Kumar et al. \cite{kumar2025using} both trained or utilized models to predict a Mean Opinion Score (MOS) as a proxy for audio quality.

\subsubsection{Reward Hacking}
Presumably a problem across all modern preference learning frameworks, reward hacking is sparsely mentioned explicitly. Step-Audio \cite{huang2025step} note their audio-to-text RL-finetuned model would exhibit hacked behavior in the form of asking for the user to repeat what they said or stating it couldn't understand audio that was clear in actuality. To mitigate this, they reject hacked responses and include rejected hacked responses in the next iteration of reward model training. Conversely, the seed-TTS RL-finetuned model \cite{anastassiou2024seed} would slow down speech and more clearly pronounce utterances at the cost of naturalness. These behaviors illustrate that optimizing against a learned reward alone can produce outputs that score well but degrade user experience, for example unnatural over articulation or strategic failure moves. This motivates the multi-signal evaluation and human calibration strategies described in the next subsection on datasets and evaluation.

\subsubsection{Multi-Modal Preference Learning}
Recent works are now attempting multi-modal applications for preference learning. The first work to do so, from Lei et al. \cite{lei2023audio}, investigated emotion recognition in speech, video, and in conjunction. Labels from different modalities were varied in weight during inference to find different modalities afforded complementary information. OpenOmni \cite{luo2025openomni} develops a multi-modal model that treats text as a pivot between image and audio modalities. DPO is only applied for emotion alignment, with audio generated with a target emotion (positive instance) and neutral emotion (negative instance). The 2 year gap between the only 2 multi-model papers in this analysis highlights the sparsity of preference-learning for multi-modal applications that include speech.

\begin{table}[b]
\centering
  \caption{Preference Dataset Segmentation}
  \label{tab:dataset_segmentation}
  \small
    \begin{tabular}{@{}l l l l p{5cm}} 
    \toprule
    \textbf{Pref. Source} &  \textbf{Ref.} \\
    \midrule
    Audio Sequential Annotation & \cite{parthasarathy2016using, parthasarathy2018preference, wu2023interval} \\
    Audio Comparison & \cite{yang2010ranking, chumbalov2020scalable, dong2020pyramid, naini2023unsupervised, naini2023preference, lopes2017modelling, cideron2024musicrl} \\
    Audio-to-Text Text Comparison & \cite{chu2024qwen2, huang2025step} \\
    Synthetic Reference Source & \cite{wu2025adaptive, zhang2024speechalign, luo2025openomni, gao2025emo} \\
    Non-Reference (Indep. Anno.) & \cite{cao2012combining, cao2015speaker, lotfian2016practical, lotfian2016retrieving, han2020ordinal, parthasarathy2017ranking, jayawardena2020ordinal, liu2021reinforcement, lei2023audio, liao2024baton, nagpal2025speech}\\
    No Preference Data (Rule-Based or Existing RM) & \cite{zhou2021interactive, anastassiou2024seed, kumar2025using} \\
    \bottomrule
  \end{tabular}
\end{table}

\begin{table}[b]
\centering
  \caption{Reference-Based Audio Datasets}
  \label{tab:datasets_distribution}
  \small
    \begin{tabular}{@{}l c l} 
    \toprule
    \textbf{Preference Source} / Dataset & \textbf{Public} & \textbf{Ref. By} \\
    \midrule
    \textbf{Audio Sequential Annotation} \\
    \quad SEMAINE Database \cite{mckeown2011semaine} & Yes & \cite{parthasarathy2016using} \\
    \quad MSP-Podcast Corpus \cite{lotfian2017building} & Yes & \cite{parthasarathy2018preference, wu2023interval, naini2023unsupervised}\\
    \quad MSP-Improv \cite{busso2016msp} & Yes & \cite{naini2023unsupervised} \\
    \addlinespace
    
    \textbf{Audio Comparison} \\
    \quad Pop Song Datasets \cite{yang2010ranking} & Yes &\cite{yang2010ranking} \\
    \quad Geographical Origin of Music \cite{zhou2014predicting} & Yes & \cite{chumbalov2020scalable} \\
    \quad MSP-Improv \cite{busso2016msp} & Yes & \cite{naini2023preference} \\
    \quad Sonancia Dataset \cite{lopes2015sonancia} & Yes & \cite{lopes2017modelling} \\
    \quad MUSHRA Audio Quality Dataset \cite{dong2020pyramid} & No & \cite{dong2020pyramid} \\
    \quad Music-RL Dataset \cite{cideron2024musicrl} & No & \cite{cideron2024musicrl} \\
    \addlinespace
    
    \textbf{Synthetic Reference Source} \\
    \quad HookTheory Dataset \cite{donahue2022melody} & No & \cite{wu2025adaptive} \\
    \quad SpeechAlign Dataset \cite{zhang2024speechalign} & No & \cite{zhang2024speechalign} \\
    \quad OpenOmni Dataset \cite{luo2025openomni} & No & \cite{luo2025openomni} \\
    \quad Emo-DPO Dataset \cite{gao2025emo} & No & \cite{gao2025emo} \\
    \bottomrule
  \end{tabular}
\end{table}

\subsection{Preference Datasets}
We synthesize the preference datasets collected or utilized in our selected works in Table \ref{tab:dataset_segmentation} and Table \ref{tab:datasets_distribution}. We first segment the means by which preferences are collected. We note that many works don't have a reference point at the time of collection but are then converted to preference pairs. Of those outlined to have a reference point and do comparisons on audio (i.e. some works compare \textit{text} generated from an audio input), we further segment and present these datasets into sequentially annotated, direct audio-to-audio comparisons, and synthetically generated reference sources. 

Sequentially annotated datasets inherently have a reference source as they refer to the last interaction to determine the current label. For example, in the process of highlighting the importance of reference-based signals when compared to independent annotation like MOS, Parthasarathy et al. \cite{parthasarathy2016using} bin emotions as decreasing, increasing, or similar to the prior 3-second bin. They apply this framework to the SEMAINE database and later the MSP-Podcast dataset \cite{parthasarathy2018preference}. Inspired by Parthasarathy, both Naini et al. \cite{naini2023unsupervised} and Wu et al. \cite{wu2023interval} apply a similar approach to the MSP-Podcast Corpus, with Naini et al. additionally applying it to the MSP-Improv dataset.

We find 6 datasets that were used to directly compare audio with a reference point, 4 of which are publicly available. The applications of these datasets are varied, with half related to music. It's worth noting that the Pop Song Datasets were published in 2010 \cite{yang2010ranking} while Music-RL in 2024 \cite{cideron2024musicrl}, displaying the persistent relevance of ranking and preference learning for music classification and generation. MSP-Improv and Pop Song Datasets were utilized for affect classification for speech and music respectively. 

Of the 4 synthetically generated preference datasets, none are publicly available. Synthetic references are usually generated by having some gold standard data that is perturbed and assumed to be worse than the original. This can either happen through randomly combining melody-chord pairs compared to the original \cite{wu2025adaptive}, targeted emotion generation (with neutral emotion serving as the negative instance) \cite{gao2025emo, luo2025openomni}, or using a stronger tokenizer \cite{zhang2024speechalign}.

\subsubsection{Synthetic Preferences}
Three distinct philosophies have emerged in synthetic preference construction. The first, exemplified by ReaLchords \cite{wu2025adaptive}, defines negative examples through random pairing, assuming that randomly matched melody-chord pairs naturally produce inferior music. The second approach, demonstrated by SpeechAlign \cite{zhang2024speechalign}, treats all synthetic audio as inherently inferior to real recordings, essentially using the generation process itself as a quality indicator. The third strategy, employed by Emo-DPO \cite{gao2025emo}, maintains semantic equivalence while varying stylistic attributes, creating negative examples through emotional mismatches rather than content corruption.

\subsubsection{Human Preferences}
Despite the prevalence of synthetic approaches, recent industry efforts have demonstrated that human preferences remain irreplaceable for achieving state-of-the-art performance. However, these same efforts have exposed troubling limitations in human evaluation of audio content. Cideron et al.'s \cite{cideron2024musicrl} finding that music generation achieves only 60\% annotator agreement, compared to 75\% for text summarization, reveals a fundamental challenge: audio quality assessment may be inherently more subjective than text.

Early works aimed to improve ranking label consistency and data adaptation quality. For example, Lotfian et al. \cite{lotfian2016practical} investigated the conversion of continuous labels to binary pairwise comparisons. They identified a margin between continuous labels to qualify data points to be included in the dataset; with larger margins resulting in higher quality labels. Following this theme, Parthasarathy et al. \cite{parthasarathy2016using} use qualitative agreement (QA) to improve preference labels. This was achieved by collecting continuous labels which were then segmented into bins and compared across annotators by establishing a threshold of agreement for inclusion. We note the need to trade off number of data points with quality of labels \cite{naini2023unsupervised}, and that smaller distances between ratings may provide valuable nuance which may require multiple iterations to recognize \cite{zhang2024speechalign}. 

Follow-up works would confirm the improvements of QA to novel architectures. QA was applied to the rankSVM architecture and a domain novel rankMargin architecture \cite{parthasarathy2018preference}, RankNet \cite{naini2023preference}, and with less granularity and rater specific considerations (i.e. relative ordinal distribution) \cite{wu2023interval}. Similarly, Lotfian et al. \cite{lotfian2016retrieving} consider a probabilistic framework that considers individual raters contribution instead of simply aggregating (i.e. mean) to form data pairs. Naini et al. \cite{naini2023unsupervised} would apply chunk-based segmentation and adversarial domain adaptation to RankNet to improve speech emotion recognition cross-corpus generalization. Despite these early works confirming the benefits of QA, no modern works use QA in their data annotation. These techniques increase label consistency but also reduce dataset size, creating a quality quantity tradeoff.

In more modern works, the subjectivity problem manifests differently. Step-Audio \cite{huang2025step} addresses it by deliberately leaving evaluation criteria (instruction following, naturalness, safety) undefined, allowing annotators to apply their own interpretations. While this approach acknowledges subjective variation, it potentially introduces inconsistency that undermines the training signal's reliability. In contrast, BATON \cite{liao2024baton} constrains evaluation to binary decisions with a "skip" option, sacrificing granularity for consistency.
The tension between human preference quality and cost has led to hybrid approaches that strategically deploy human annotation. Cideron et al.'s \cite{cideron2024musicrl} multi-reward framework uses human preferences only for overall quality assessment while relying on automated metrics for specific attributes like text adherence. Additionally, longer audio segments led to higher inter-annotator agreement, at least in SER tasks when comparing 1, 3, and 5 second segments \cite{wu2023interval}, hinting at the need for more human-centered evaluation approaches. This selective use of human judgment suggests the field is moving toward a nuanced understanding of where human evaluation adds unique value versus where it introduces unnecessary noise.

\begin{table}[b]
\centering
  \caption{Evaluation Strategies}
  \label{tab:evaluation_distribution}
  \small
    \begin{tabular}{@{}l c c} 
    \toprule
    \textbf{Evaluation} &\textbf{Subj./Obj.} & \textbf{\# Used} \\
    \midrule
    Error Metrics (ACC, UAR, P@K, RMSE, EER, MAE) & Obj. & 14 \\ 
    Correlation Metrics (Kendall's, Spearman, Pearson's, Goodman-Kruskal) & Obj. & 11 \\ 
    Content Accuracy (WER, CER, SIM) & Obj. & 7 \\ 
    LLM Evaluation & Obj. & 3 \\ 
    Distribution Metrics (KL vs Reward, KL, Frechet, KS-Test) & Obj. & 3 \\ 
    Distance Metrics (Cross-Entropy, SDTW, Quantization Error) & Obj. & 3 \\ 
    Perceptual Metrics (Emotion Change Consensus, CMLAP, Inception Score) & Obj. & 2 \\ 
    Audio Quality (PESQ, SSNR, SI-SDR, STOI) & Obj. & 1 \\ 
    Musicality (Harmonic Quality, Synchronization, Rhythmic Diversity) & Obj. & 1 \\ 
    \addlinespace
    Human Mean Opinion Score & Subj. & 11 \\ 
    Human A/B Comparison & Subj. & 2 \\ 
    HCI (Semi-Structured Interview/Creative Support Index) & Subj. & 1 \\ 
    \bottomrule
  \end{tabular}
\end{table}

\subsection{Evaluation}
We synthesize the highlighted work's evaluation metrics in Table \ref{tab:evaluation_distribution}; for a full breakdown of these metrics by paper see Table~\ref{tab:full_evaluation_distribution} in the Appendix. The line between evaluation and reward metrics blurs in this context, as metrics used to assess final performance often overlap with those guiding the learning process. Pre-modern preference learning frameworks often use error or correlation metrics for their classification tasks. Accuracy and Precision at K are particularly used to evaluate classification or ranking models. More domain-specific metrics, such as Perceptual Evaluation of Speech Quality (PESQ) and Harmonic Quality for music see some use as well by one-off papers. Speech recognition tasks utilize Word Error Rate (WER) and the more fine-grained Character Error Rate (CER), albeit more sparsely. Some of these metrics we classify as objective, but contain more nuance as their source may be from human labels (e.g. NISQA). PESQ is a model of human perception can could be considered subjective. Similarly, LLM evaluations could also be construed as subjective but we view pretrained models as objective in this work.

\subsubsection{Alignment Between Objective and Subjective Metrics}
Many works, typically related to generative tasks, use human evaluations to evaluate their models. These types of evaluations, especially with a large number of annotations, is viewed as the gold standard. 13 of the 30 works use subjective metrics such as Mean Opinion score and A/B comparisons, both of which seems to have wide applications from quality to faithfulness to musicality. This type of evaluation highlights the difficulties of evaluating generative audio systems and how automated metrics may not be enough. 

Since human evaluations are considered a gold standard, some of these works inform the alignment between objective, easy to collect, metrics with more expensive subjective metrics. 12 works used both subjective and objective metrics in the evaluation of their system and we synthesize 9 relevant works in Table \ref{tab:eval_alignment}. Despite error metrics being the most used form of evaluation, multiple works \cite{yang2010ranking, kumar2025using} highlight the misalignment between error metrics like mean squared error (MSE) with subjective evaluations like music ranking satisfaction \cite{yang2010ranking} and quality estimation \cite{kumar2025using}.

\begin{table}[h]
\centering
    \caption{Alignment Between Automatic and Human Scores}
    \label{tab:eval_alignment}
    \footnotesize
        \begin{tabular}{@{}llclc@{}}
        \toprule
        \textbf{Task} & \textbf{Metric} & \textbf{Align} & \textbf{Human Evaluation} & \textbf{Ref.} \\
        \midrule
        \multicolumn{5}{@{}l}{\textit{Recognition Tasks}} \\
        Music Emotion & MSE & $\times$ & Ranking Satisfaction (0-10) & \cite{yang2010ranking} \\
        Speech (ASR) & WER & $\times$ & SLP Error Severity (1-3) & \cite{nagpal2025speech} \\
         & MPS & \checkmark & SLP Error Severity (1-3) & \cite{nagpal2025speech} \\
        Speech Quality & MSE & $\times$ & NISQA (1-5) & \cite{kumar2025using} \\
         & SSNR, SI-SDR & \checkmark & NISQA (1-5) & \cite{kumar2025using} \\
        \addlinespace
        
        \multicolumn{5}{@{}l}{\textit{Generation Tasks}} \\
        Speech Emotion & WER, P-SIM, SER & \checkmark & Quality/Emotion MOS (1-5) & \cite{gao2025emo} \\
         & WER, SIM & \checkmark & Naturalness/Faithfulness A/B & \cite{zhang2024speechalign} \\
         & WER & ? & Similarity/Expressiveness CMOS (-2,+2) & \cite{anastassiou2024seed} \\
         & SIM & $\times$ & Similarity/Expressiveness CMOS (-2,+2) & \cite{anastassiou2024seed} \\
        Music & MuLAN & $\times$ & CMOS (1-5) & \cite{cideron2024musicrl} \\
         & Audio Quality & \checkmark & CMOS (1-5) & \cite{cideron2024musicrl} \\
         & Melody-Chord Comp. & \checkmark & Musicality CMOS (1-5) & \cite{wu2025adaptive} \\
         & Melody-Chord Discr. & \checkmark & Musicality CMOS (1-5) & \cite{wu2025adaptive} \\
        Text-to-Audio & CLAP & \checkmark & Faithfulness MOS (1-5) & \cite{liao2024baton} \\
        \bottomrule
        \end{tabular}
        
        \vspace{0.5em}
        \small
        \checkmark = Aligned, $\times$ = Misaligned, ? = Unclear. SLP = Speech-Language Pathologist.
\end{table}

\subsubsection{Alignment in Speech}
Word error rate (WER) presents a particularly nuanced picture in its relationship with human judgment. The literature reveals contradictory findings: some studies demonstrate alignment between WER and subjective scores related to quality, naturalness, and faithfulness \cite{zhang2024speechalign, gao2025emo}, while others find it detrimental to disordered speech recognition \cite{nagpal2025speech}. Interestingly, \cite{anastassiou2024seed} suggest an optimal point exists where WER approaches that of human speech error rate. This variability in findings suggests that the relationship between WER and subjective quality may be context-dependent, requiring careful consideration in reward signal or evaluation frameworks.

Several metrics show stronger correlation with human judgments, offering more reliable alternatives for automated evaluation. The Meaning Preservation Score (MPS) demonstrates correlation with subjective scores in disordered speech recognition \cite{nagpal2025speech} while SSNR and SI-SDR scores exhibit high correlation with NISQA scores based on subjective perception of quality \cite{kumar2025using}. These findings suggest that metrics designed to capture semantic content or overall quality patterns may better approximate human perception than traditional error-based measures.

Speaker similarity metrics reveal complex alignment patterns with human judgments. While both SIM and prosody similarity (P-SIM) are used to evaluate speech emotion generation, their correlation with subjective assessments varies significantly by evaluation criteria. P-SIM demonstrates alignment with quality/naturalness and emotion similarity scores \cite{gao2025emo}, while standard SIM aligns with naturalness/faithfulness ratings \cite{zhang2024speechalign}. However, SIM shows misalignment with similarity/expressiveness evaluations \cite{anastassiou2024seed}. This divergence suggests that alignment may depend more on what aspect of quality is being evaluated than on the metric itself. The prosody-specific variant (P-SIM) may better capture emotional nuance, explaining its stronger correlation with emotion-related subjective scores, while general speaker similarity may be sufficient for faithfulness but inadequate for capturing expressive qualities.

\subsubsection{Alignment in Music}
In musical applications, the alignment between objective and subjective metrics reveals additional challenges. While audio quality showed only marginal improvement relative to overall MOS scores \cite{cideron2024musicrl}, the MuLan score actually decreased when MOS increased, indicating a potential trade-off between different quality aspects. Wu et al. \cite{wu2025adaptive} found that Melody-Chord compatibility and Melody-Chord Discrimination showed some relationship with musicality-based subjective scores. Cideron et al. \cite{cideron2024musicrl} explicitly note the lack of correlation between audio quality and text-to-music prompt adherence in subjective holistic evaluations. 

\subsubsection{Critical Gaps}
This variance in metric-human alignment patterns highlights a critical gap: while many works employ ensemble or specialized metrics, few explicitly investigate how these signals relate to one another, to holistic quality assessments, or their susceptibility to reward hacking. The relationship between targeted subjective metrics (e.g., musicality ratings) and overall preference judgments (e.g., A/B comparisons, CMOS) remains largely unexplored, despite the field's increasing reliance on multi-metric and subjective evaluation frameworks.

\section{Discussion}
\label{sect:discussion}
In the previous section we systematically reviewed and synthesized our findings from 30 works spanning 2010-2025, revealing both the current state and future potential of preference learning in audio.  We now examine why audio remains underexplored despite a clear need (Section \ref{sec:current_state}), and then analyze emerging methodological patterns including evaluation strategies, preference generation approaches, and algorithmic considerations (Section \ref{sec:insights}). Lastly, we discuss gaps and future directions for the field (Section \ref{sec:gaps}). Overall our work reveals a field in transition, with promising techniques from pre-modern works yet to be integrated into modern frameworks.

\subsection{Current State of Preference Learning in Audio} \label{sec:current_state}
Despite recent growth—with 9 audio generation studies emerging in just 5 years—preference learning in audio remains remarkably underexplored. Of approximately 500 manually reviewed works, despite a focused keyword search only 30 (6\%) apply preference learning to audio applications while 157 (30\%) apply preference learning to other domains, This scarcity is particularly striking given audio's parallel challenges with NLP in evaluation metrics. We posit that this is due to the lack of dataset availability, proprietary considerations, and compounding of resource costs that are already fairly large in the text domain. 

Our analysis reveals that the field is undergoing a fundamental shift. Pre-2021 works focused primarily on classification tasks using traditional methods like rankSVM for emotion recognition. Post-2021, we observe a dramatic pivot toward generation tasks employing modern RLHF/DPO frameworks. Three critical patterns in methodology emerge: 

\begin{enumerate}
    \item Multi-dimensional evaluation strategies, with synthetic and automated preferences proving effective for specific needs while human judgments remain essential for holistic evaluation
    
    \item Inconsistent alignment between traditional metrics and human judgments across different contexts, necessitating strategic metric selection in reward signal design
    
    \item Convergence on multi-stage training pipelines that strategically combine different reward signals
\end{enumerate}

\subsection{Notable Insights} 
\label{sec:insights}

\subsubsection{Multi-Dimensional Evaluation Strategies}
We identify four dimensions that may be applicable depending on the task and context: content preservation, production quality, task achievement, and human preference. Content preservation thus far leverages semantic metrics such as WER, CER, and MPS in addition to speaker identity metrics like SIM or P-SIM. Content preservation acts almost like a constraint to maintain either semantic or prosody similarities while considering other task specific optimizations like emotion expressivity. For example, you might want to ensure your SIM score stays high while finetuning your model to have emotion. This prevents a general shift away from initial intonation as other types of voices might naturally sounds more angry or sad.

Task achievement metrics noted in this corpus optimize emotion expressivity, musicality, and instruction following to date. SER classification and emotion MOS can be included to improve emotion expressivity of speech and potentially music. Harmony and melody-chord alignment signals can improve overall musicality of songs, and CLAP embeddings can indicate alignment of generated audio to intended textual input. These are typically specific signals that while weaker than human preferences and lacking in nuance, provide a nice starting point to optimize to if human preferences are unavailable. Each may come with its own assumptions and biases that need to be addressed through the inclusion of another signal. For example, optimizing solely for SER classification accuracy might produce emotionally exaggerated speech that sounds unnatural, necessitating balancing with naturalness metrics.

Production quality encompasses both objective signal characteristics and subjective naturalness. Traditional metrics show fundamental limitations: PESQ, originally designed for degraded telephony, demonstrates poor correlation with modern neural vocoder quality \cite{manocha2020differentiable}. Our analysis reveals context-dependent performance—while SSNR and SI-SDR correlate strongly with subjective NISQA scores \cite{kumar2025using}, these same metrics fail to capture expressive qualities. Critical trade-offs emerge in practice: seed-TTS \cite{anastassiou2024seed} slowed speech and over-articulated pronunciations to improve WER at the expense of naturalness, while Step-Audio \cite{huang2025step} documented reward hacking where models would claim inability to understand clear audio. This tension between objective improvements and perceptual quality underscores why 13 works employ both objective and subjective metrics, acknowledging that no single metric captures the multifaceted nature of audio quality.

Human preference's strengths lie in their holistic evaluation and flexibility. Holistic, ill-defined prompts of overall quality leverage subjectivity to develop a more generalizable end model but may demand more data points. Task-specific or clearly defined prompts in human preference solicitation remove some of this subjectivity and eventual generalization in favor of a more interpretable streamlined signal. In this corpus, multiple studies look towards an overall signal of musicality \cite{wu2025adaptive}, naturalness \cite{anastassiou2024seed, zhang2024speechalign}, or general quality \cite{cideron2024musicrl}, while others employ streamlined signals through frameworks like MUSHRA \cite{dong2020pyramid}, which provides controlled comparison of audio quality against hidden references and anchors. 

These four dimensions are fundamentally interdependent: optimizing one often degrades others. As demonstrated earlier, pursuing task achievement alone (e.g., maximizing SER accuracy) can compromise production quality (naturalness), while improving content preservation (lowering WER) may require trade-offs in expressiveness. This interdependence necessitates multi-objective evaluation frameworks that balance competing priorities rather than optimizing single metrics in isolation.

We recommend a staged approach: begin with task achievement signals to establish basic functional competence, then introduce content preservation and production quality constraints to prevent degradation of core attributes. Human preferences should serve as either (1) a holistic replacement when sufficient data is available (requiring more annotations but less reward balancing), or (2) targeted complementary signals that address specific gaps left by automated metrics (requiring strategic collection but enabling more interpretable debugging of model behavior). This staged approach allows early detection of reward hacking before expensive human annotation. Notably, our review suggests a critical threshold around 50k preference pairs \cite{zhang2024speechalign}, beyond which data quality improvements outweigh raw quantity gains—a finding that should inform collection strategy decisions.

\subsubsection{Human vs Synthetic Preferences}
As seen in natural language processing, human evaluations are not without their own biases. In text, human's tend to prefer longer texts over shorter ones despite semantic content \cite{singhal2023long}. No works have investigated the relationship between audio length and preferences in audio although it is noted agreement rates to to be positively correlated with audio length \cite{wu2023interval}, at least for the case of emotion change considering 1, 3, and 5 seconds intervals. While human preferences may be biased, they often contain much more nuance than synthetic preferences.

We identify three methods of synthetic preference pair generations. Random pairing, in which data is segmented into two ideally complementary components, like melody and chord accompaniment \cite{wu2025adaptive}, before randomly pairing data points and treating the original "real" pair as the preferred instance. This may be noisy as some random pairings may actually go fairly well together. Those that clearly do not pair well may be easily discriminated against, leading to a weak reward model. 

Additionally, training two generative models on data of different qualities—such as one vocoder trained on real speech and another trained on synthetic data \cite{zhang2024speechalign}—could create assumed real/synthetic pairs where the real data consistently produces superior results. This method offers reliable ground truth for quality comparisons but may introduce systematic biases related to the specific training data distributions rather than perceptual quality differences alone.

Finally, several studies use conditional generation to produce both target outputs and non-target alternatives, treating the target as preferred and all others as non-preferred. This approach is most common in speech synthesis, where researchers generate audio with specific emotional qualities. For example, when the target emotion is sadness, the system generates speech that sounds sad (preferred) alongside speech expressing other emotions (non-preferred) \cite{gao2025emo, luo2025openomni}. However, this approach lacks nuance and may oversimplify the task, though it could be particularly useful for distinguishing high arousal emotions that are commonly misclassified \cite{al2023speech}. Synthetic preferences remain a practical substitute for costly human annotation but risk overfitting to structural biases in training data rather than perceptual quality. Hybrid approaches that combine synthetic signals with periodic human calibration could mitigate this limitation.

\subsubsection{Data Collection Strategies}
Modern works reveal persistently low inter-annotator agreement in audio tasks, particularly for music generation where agreement rates are notably lower than in NLP \cite{cideron2024musicrl}. While pre-modern works developed methods to improve agreement in emotion and speech recognition tasks through quality assurance frameworks, these techniques have yet to be systematically applied to or acknowledged by modern preference learning generative audio applications. Specifically, thresholds of agreement between annotators might be considered. 

Some works found lower threshold agreements were sufficient for easier tasks like SER of arousal, valence, and expectation whereas power improved when agreement was at 100\% \cite{parthasarathy2016using}. Qualitative agreement, which considers agreement relative to annotators past annotations (i.e. this is clip is happier than the last clip in this sequence) proved effective for SER task improvement \cite{parthasarathy2018preference}. Additionally, thresholds on the distance between preferences scores may be implemented. On a 5 point scale, still a SER task, a margin of 1 and 2 performed well while a margin greater than that decrease model performance \cite{parthasarathy2017ranking}. Though, larger differences between annotations made data more reliable \cite{lotfian2016practical}. In summary, depending on the difficulty of a task you may filter your preference pairs by annotator agreement, relative agreement (QA), or by scoring difference. Each likely improves the quality of labels though at the cost of number of data points.

As for the number of instances needed for effective fine-tuning, some works indicate that 50k pairs was sufficient and saw no increase relative to 250k pairs of instances \cite{zhang2024speechalign}. This was shown specifically for in speech model alignment using DPO. Zhang et al. highlight that once the 50k threshold is met, quality of data becomes of more importance than dataset size. In summary, high-quality annotation frameworks through inter-rater filtering, relative scoring, or margin-based selection could outperform by purely scaling data quantity

\subsubsection{Reward Signal Design}
Synthetic or automated signals are particularly useful when desired fine-tuning is along a simple axis, like emotion type or speaker identity, where the distinction between preferred and non-preferred instances is relatively clear-cut. Human preferences allow for more nuance, specially when instances being compared are fairly similar to one another. Given the costs associated with human preferences, synthetic signals used in combination with automated or other synthetic signals can provide a strong alternative approach but require careful balancing. For example, while audio quality may be important at first, eventual generative models may need to develop better text adherence once the model can consistently generate high quality audio.

Using multiple reward signals does also have the benefit of interpretability. While needing to balance tradeoffs between signals can make training unstable, it paints a clearer picture of what is actually being optimizing. Comparatively, holistic human/automated evaluations may have nuance but could also contains undesired biases to be exploited via reward hacking, like the aforementioned response length which we anticipate may impact audio as well.

While sparsely mentioned in these works, some reward hacking mitigation methods are worth noting. KL-Divergence is used in virtually all modern works to prevent deviation from a reference policy. Particularly in Step-Audio \cite{huang2025step}, when reward hacking was observed, the hacked responses were included in the following training iteration as rejected responses. This approach would likely generalize to other forms of reward hacking. We also note that reward signals need not emphasize simply improving scores, i.e. lower WER, but may aim to bring signals closer to human performances on scores. In the case of speech synthesis, a WER closer to human WER sounded more natural than simply lowering WER when optimizing \cite{anastassiou2024seed}. 

\subsubsection{Algorithmic Considerations}
Multiple works have shown the utility multiple preference learning algorithms including REINFORCE, PPO, and DPO predominantly on transformer architectures \cite{wu2025adaptive, huang2025step, luo2025openomni}. Notable exceptions apply a MSE reward framework to a diffusion model \cite{liao2024baton}. Results show general cross-task performance gains, possibly due to positive publication biases, when compared to a supervised models. Explicitly, additional SFT training steps didn't account for preference learning improvements \cite{zhang2024speechalign}. Preference learning frameworks also tended to make performance more generalizable to unseen datasets relative to SFT baselines for SER \cite{naini2023preference}. For classification and generative tasks, preference learning frameworks likely apply well to the audio domain across algorithms and architectures. 

DPO is perhaps the most convenient training framework as it bypasses the need for a computationally expensive external reward model. External reward models can be seen as more flexible since multiple reward signals can be incorporated into the training framework, but MO-DPO may enable multi-objective training  \cite{zhou2023beyond}. Of particular importance, on single iteration tasks PPO seems to perform best relative to a single iteration of DPO, at least on speech alignment tasks \cite{zhang2024speechalign}. We highlight the finding, as it relates to DPO given high quality data, multiple iterations may seem increasing gains after the one training iteration. This encourages the collection of high quality nuanced data that may not lead to immediate but eventual performance improvements, similarly to a form of curriculum learning. Preference-based optimization yields reliable gains across architectures when supported by sufficient data diversity, suggesting algorithmic choice may matter less than data quality in determining downstream success.

\subsection{Gaps and Future Directions}
\label{sec:gaps}
\subsubsection{Underexplored Applications}
While emotion is a strong starting point, and speaks to the strength of preference learning's ability to learn from subjectivity, additional applications could leverage this framework. Particularly, more specific evaluations in music and prosody considerations beyond emotion could be considered. For example, evaluations on how musical phrasing aligns with compositional intent, or how speech prosody conveys pragmatic meaning like sarcasm or emphasis. Preference learning could extend to capturing subtler musical qualities like groove, swing, or stylistic authenticity that are difficult to quantify but readily perceived by listeners. In relation to audio quality, preference learning has additional applications. Active noise canceling, general audio restoration, and source separation could all benefit from measures of subjectivity. 

Furthermore, multi-modal integrations of preference learning need to be further explored. While recent works beyond the time frame of this survey explore multi-modal preference integration \cite{xu2025qwen2, hurst2024gpt, team2024gemini}, the proprietary nature of these systems leaves critical questions unaddressed. Specifically, whether audio-specific temporal factors (discussed in Section \ref{Technical_Challenges}) transfer to multi-modal contexts, and how to design annotation protocols that capture cross-modal quality trade-offs, remain open problems. Audio's sequential nature and lower inter-annotator agreement suggest that multi-modal preference annotation requires fundamentally different protocols than text-only approaches, capturing not only within-modality quality but also cross-modal coherence and interaction effects.

\subsubsection{Evaluation and Benchmarking Needs}
Cross-tasks, while there are some attempts at developing benchmarks, there is no consensus on what the benchmark should be. The MSP-Podcast Corpus \cite{lotfian2017building} is the most widely used speech dataset that could serve as a starting point for speech emotion evaluations. Two works include their own evaluation frameworks that could be adopted. Namely, StepEval-Audio-360 \cite{huang2025step} and OmniBench \cite{li2024omnibench} are used. StepEval-Audio-360 provide a benchmark that evaluates multi-modal speech. OmniBench focuses more on cross-modality evaluations, also including visual components in its evaluation. Which is better, for what task, and is more utilized remains to be seen. Future work is needed to establish which evaluation framework best serves different audio preference learning tasks. Future efforts could standardize evaluation protocols by defining a minimal set of shared metrics and listener test formats to ensure reproducibility. A community benchmark similar to HELM \cite{liang2022holistic} for text or SUPERB \cite{yang2021superb} for speech could unify disparate evaluation practices.

\subsubsection{Technical Challenges Unique to Audio}
\label{Technical_Challenges}
The lower agreement rate in the audio domain is problematic for stable training. Future work may consider leveraging pre-modern frameworks to improve the agreement rates via inter-rater agreement thresholds, qualitative agreement, and point wise difference margins. Additionally, work should investigate the impact of audio length of agreement beyond 5 second intervals, as it may improve agreement rates. Other considerations from Human-Computer Interaction (HCI) to improve the interface presented to users could be of use as well. 

The temporal nature of audio presents unique challenges and opportunities for preference learning. Unlike text, which can be skimmed or reviewed holistically, audio unfolds sequentially, requiring real-time processing and preventing reviewers from easily accessing later content before making judgments. This sequential constraint may enable more controlled evaluation conditions but could also introduce recency bias. Future work should investigate how playback controls (fast-forwarding, replay) affect inter-annotator agreement and whether continuous, time-aligned preference annotations could capture granular feedback without disrupting the listening experience.

Furthermore, audio preference learning faces unique challenges from cultural and linguistic variation. Accent, prosody, and speaking style immediately signal regional and social backgrounds, affecting preference judgments in ways that text does not. Emotional expressions deemed appropriate vary dramatically across cultures—what sounds professional or emotionally appropriate in one context may be perceived differently in another. Our review reveals these challenges are largely unaddressed, with most studies limited to English or Mandarin. This cultural dependency suggests preference models trained on one population may not generalize, necessitating either culturally-diverse training data or explicit modeling of cultural context. Future work should investigate whether preference learning can identify and mitigate these biases or whether separate models are needed for different cultural contexts. This might be accomplished by incorporating multilingual and culturally diverse datasets to ensure generalizable preference alignment across listening populations.




\section{Limitations}
\label{sect:limitations}
While the analytic and interpretive components of the review were conducted jointly by the authors, the screening and eligibility assessments were performed by a single reviewer. Although our predefined protocol still promoted internal consistency, the absence of a second reviewer prevents measurement of inter-rater reliability and increases the risk of selection bias. Future updates to this survey would benefit from an independent duplicate screening step to validate inclusion decisions.

We screened arXiv papers by top 20\% most-cited, treating relative citation count as a proxy for paper quality. While citation count serves as a useful heuristic, this approach inherently biases against more recent submissions that have had limited time to accumulate citations and against less-popular research topics. We acknowledge this limitation, but given the infeasibility of manually reviewing all 1,627 potential arXiv papers, this method offers a practical and scalable compromise.

Our analysis reveals a geographic and linguistic bias, with most studies originating from the United States and focusing on English-language content. While some studies include Mandarin (using character error rate as the counterpart to word error rate) and German speech emotion recognition, few investigate how cultural and accent variations impact performance. This Western-centric focus may limit the generalizability of our findings to global audio applications, particularly given that emotion recognition accuracy may vary across cultures and speech recognition can be impacted by accents.

\section{Conclusion}
\label{sect:conclusion}
This survey provides the first systematic examination of preference-based learning in the audio domain, identifying that only 6\% of roughly 500 screened papers directly leverage preference signals in this domain. Despite the rapid growth of preference-learning methods in NLP, our analysis shows that the audio community has adopted these techniques unevenly, with substantial variation in annotation protocols, data-collection strategies, and reward-modeling pipelines. This fragmentation, combined with persistent challenges such as annotation cost, reward hacking, and unstable optimization, highlights the need for greater methodological cohesion in future work.

Across the literature, we find limited alignment between commonly used audio metrics and human perceptual judgments, underscoring the necessity of evaluation schemes that are explicitly grounded in listener preferences—rather than inherited from text or signal-processing traditions. Consistent benchmarks, high-quality preference datasets, and multimodal or ensemble-feedback signals represent promising directions to close this gap. In particular, lessons from large-scale RLHF pipelines in NLP suggest that scalable preference collection, reliable reward modeling, and robust regularization strategies will be essential for building audio systems that better capture subjective human perception.

By organizing existing efforts and revealing where evidence converges or diverges, this review aims to serve as a foundation for future research in preference-based audio modeling. As generative audio systems become more capable and more widely deployed, developing principled and human-aligned preference-learning frameworks will be critical for ensuring that audio models faithfully reflect the nuanced judgments of their listeners.


\begin{acks}
Funding for this project was supported by the NIH National Library of Medicine's T15 Biomedical Informatics and Data Science Research Training Program (Grant T15LM011271). 
\end{acks}

\bibliographystyle{ACM-Reference-Format}
\bibliography{references}

\newpage

\appendix
\counterwithin{figure}{section}
\counterwithin{table}{section}
\section{Appendix}

\subsection{Screening Outcomes Across Sources}
Table~\ref{tab:screening_reasons} summarizes how papers from different sources were classified during screening, illustrating the distribution of works excluded for lacking either audio relevance or preference-learning components.

\begin{table}[h]
  \centering
  \caption{Reasons for Removal of Papers from Corpus}
  \label{tab:screening_reasons}
  \vspace{-1em}
  \begin{tabular}{@{} l l l @{}}
    \toprule
    \parbox[t]{4.5cm}{\textbf{Source} \\ Included/Removal Reason} & \parbox[t]{1cm}{\textbf{Total}\\Count} &  \parbox[t]{1.8cm}{Count/\textbf{Total}} \\
    \midrule
    \textbf{Top Venues} & \textbf{116} & \textbf{100.00\%} \\
    Included (Both Audio + PL) & 10 & 8.62\% \\
    Audio only (not PL) & 12 & 10.34\% \\
    PL only (not Audio) & 46 & 39.66\%  \\
    Neither Audio nor PL & 48 & 41.38\% \\
    Survey & 0 & 0.00\% \\ 

    \addlinespace

    \textbf{arXiv} & \textbf{345} & \textbf{100.00\%} \\
    Included (Both Audio + PL) & 7 & 2.03\% \\
    Audio only (not PL) & 17 & 4.93\% \\ 
    PL only (not Audio) & 99 & 28.69\% \\ 
    Neither Audio nor PL & 132 & 38.26\% \\ 
    Survey & 90 & 26.09\% \\ 
    \addlinespace

    \textbf{Relevant Citations} & \textbf{60} & \textbf{100.00\%} \\    
    Included (Both Audio + PL) & 13 & 21.67\% \\
    Audio only (not PL) & 31 & 51.67\% \\ 
    PL only (not Audio) & 12 & 20.00\% \\ 
    Neither Audio nor PL & 2 & 3.33\% \\
    Survey & 2 & 3.33\% \\ 
    \addlinespace

    \textbf{All Sources} & \textbf{521} & \textbf{100.00\%} \\    
    Included (Both Audio + PL) & 30 & 5.76\% \\
    Audio only (not PL) & 60 & 11.52\% \\ 
    PL only (not Audio) & 157 & 30.13\% \\ 
    Neither Audio nor PL & 182 & 34.93\% \\ 
    Survey & 92 & 17.66\% \\ 
    \bottomrule
  \end{tabular}
\end{table}

\vspace{-1em}
\subsection{Citation-Based arXiv Sampling}
Table~\ref{tab:top20-screening} reports the screening results for the top 20\% most-cited arXiv papers per year, showing how citation-driven sampling contributed to the final set of included works.

\vspace{-.5em}
\begin{table}[h]
  \centering
  \caption{Screening of the top 20\% most-cited arXiv papers by year}
    \vspace{-1em}
  \label{tab:top20-screening}
  \begin{tabular}{@{}lrrrrr@{}}
    \toprule
    \textbf{Year} & \textbf{Total papers} & \textbf{Citation cutoff} & \textbf{Reviewed} & \textbf{Included} \\
    \midrule
    2020 &  19 & $\geq$ 28 &   4 & 1 \\
    2021 &  27 & $\geq$ 22 &   6 & 0 \\
    2022 &  37 & $\geq$ 32 &   8 & 0 \\
    2023 & 337 & $\geq$ 90 &  68 & 1 \\
    2024 & 926 & $\geq$ 13 & 195 & 5 \\
    2025 & 291 & $\geq$  1 &  64 & 2 \\
    \bottomrule
  \end{tabular}
  \vspace{-1em}
\end{table}

\subsection{Temporal Distribution of Included Works}
Table~\ref{tab:year_dist} provides the publication year distribution of all included papers, highlighting the recent acceleration of preference-learning work in audio.

  \vspace{-1em}
\begin{table}[h]
\centering
  \caption{Distribution of Selected Works by Year}
    \vspace{-.5em}
  \label{tab:year_dist}
  \small
  \begin{tabular}{@{}c l l p{6cm}@{}}
    \toprule
    \textbf{Year} & \textbf{Ref.} \\
    \midrule
    2010 & \cite{yang2010ranking} \\
    2011 & \\
    2012 & \cite{cao2012combining} \\
    2013 & \\
    2014 & \\
    2015 & \cite{cao2015speaker} \\
    2016 & \cite{lotfian2016practical, lotfian2016retrieving, parthasarathy2016using} \\
    2017 & \cite{parthasarathy2017ranking, lopes2017modelling} \\
    2018 & \cite{parthasarathy2018preference} \\
    2019 & \\
    2020 & \cite{han2020ordinal, chumbalov2020scalable, dong2020pyramid, jayawardena2020ordinal} \\
    2021 & \cite{zhou2021interactive, liu2021reinforcement} \\
    2022 & \\
    2023 & \cite{naini2023unsupervised, naini2023preference, wu2023interval, lei2023audio} \\
    2024 & \cite{cideron2024musicrl, anastassiou2024seed, chu2024qwen2, liao2024baton} \\
    2025 & \cite{nagpal2025speech, wu2025adaptive, huang2025step, luo2025openomni, gao2025emo} \\
    \bottomrule
  \end{tabular}
\end{table}

\subsection{Emotion Representations Used in Prior Work}
Table~\ref{tab:emotions_considered_distribution} catalogs the different emotion representations—dimensional, categorical, and hybrid—used in the literature, illustrating the diversity of labeling schemes across tasks.

\begin{table}[h]
\centering
  \caption{Emotion Representations in Audio Preference Learning Tasks.}
    \vspace{-.5em}
  \label{tab:emotions_considered_distribution}
  \small
    \begin{tabular}{@{}l p{7cm} c c} 
    \toprule
    \textbf{Task/Approach} & \textbf{Emotion Representation} & \textbf{Works} & \textbf{Count}\\
    \midrule
    \textit{Dimensional Approaches} \\
    Music Emotion Rec. & Valence-Arousal (VA) & \cite{yang2010ranking} & 1 \\
    Soundscape Emotion Rec. & Valence-Arousal-Tension (VAT) & \cite{lopes2017modelling} & 1 \\
    Speech Emotion Rec. & Valence-Arousal (VA) & \cite{lotfian2016practical} & 1 \\
     & Valence-Arousal-Dominance (VAD) & \cite{parthasarathy2017ranking, parthasarathy2018preference, naini2023unsupervised, naini2023preference} & 4 \\
     & Valence-Arousal-Power-Expectation (VAPE) & \cite{parthasarathy2016using} & 1 \\
    \addlinespace
    
    \textit{Categorical Approaches} \\
    Speech Emotion Rec. & Basic emotions (4-5 categories) & \cite{lotfian2016retrieving, cao2012combining} & 2 \\
     & Basic emotions (6 categories) & \cite{lei2023audio, cao2015speaker} & 2 \\
     & Polarity-based (3 levels) & \cite{han2020ordinal} & 1 \\
    Speech Emotion Gen. & Basic emotions (4 categories) & \cite{anastassiou2024seed, liu2021reinforcement} & 2 \\
     & Basic emotions (5 categories) & \cite{gao2025emo} & 1 \\
     & Basic emotions (6 categories) & \cite{luo2025openomni} & 1 \\
    \addlinespace
    
    \parbox[t]{4cm}{\textit{Hybrid Approaches\\(Dimensional + Categorical)}} \\
    Speech Emotion Rec. & VAD + 7 discrete emotions & \cite{wu2023interval} & 1 \\
     & VAD + 12 discrete emotions & \cite{jayawardena2020ordinal} & 1 \\
    \bottomrule
  \end{tabular}
  \vspace{0.5em}
  \noindent\small\textit{Note: Dimensional approaches use continuous scales (e.g., valence-arousal), while categorical approaches use discrete emotion labels.}
\end{table}

\subsection{Evaluation Metrics Identified in the Corpus}
Table~\ref{tab:full_evaluation_distribution} summarizes the evaluation strategies used across the surveyed works, spanning objective signal-based metrics, correlation and error measures, distributional tests, and subjective human-judgment methods.

\begin{table}[h]
\centering
  \caption{Granular Representation of Evaluation Strategies}
      \vspace{-.5em}
  \label{tab:full_evaluation_distribution}
  \small
    \renewcommand{\arraystretch}{0.9}
    \begin{tabular}{@{}l c l} 
    \toprule
    \textbf{Evaluation} &\textbf{Subjective/Objective} & \textbf{Used By} \\
    \midrule
    \textbf{Perceptual Quality} \\
     Perceptual Evaluation of Speech Quality & Obj. & \cite{kumar2025using} \\
     Segmental Signal-to-Noise Ratio & Obj. & \cite{kumar2025using} \\
     Scale-Invariant Signal-to-Distortion Ratio & Obj. & \cite{kumar2025using} \\

    \textbf{Correlation Metrics} \\
     Kendall's Tao Rank Correlation & Obj. & \cite{lotfian2016retrieving, parthasarathy2017ranking, parthasarathy2018preference, lopes2017modelling, jayawardena2020ordinal, naini2023unsupervised, naini2023preference} \\
     Spearman Rank Correlation & Obj. & \cite{yang2010ranking, dong2020pyramid, lei2023audio} \\
     Goodman-Kruskal Gamma & Obj. & \cite{yang2010ranking, chu2024qwen2} \\
     Pearson's Correlation & Obj. & \cite{dong2020pyramid} \\
         
    \textbf{Error Metrics} \\
     Accuracy/UAR & Obj. & \cite{yang2010ranking, cao2012combining, cao2015speaker, parthasarathy2016using, lopes2017modelling, han2020ordinal, liu2021reinforcement, naini2023unsupervised, lei2023audio, luo2025openomni} \\
     Precision at K & Obj. & \cite{cao2015speaker, lotfian2016practical, lotfian2016retrieving, parthasarathy2017ranking, naini2023unsupervised, lei2023audio} \\
     Root Mean Squared Error & Obj. & \cite{han2020ordinal, dong2020pyramid} \\
     Equal Error Rate & Obj. & \cite{han2020ordinal} \\
     Mean Absolute Error & Obj. & \cite{dong2020pyramid} \\
     Cross Entropy & Obj. & \cite{nagpal2025speech} \\
     Soft Dynamic Time Warping & Obj. & \cite{zhou2021interactive} \\
     Quantization Error & Obj. & \cite{wu2023interval} \\
         
    \textbf{Distribution Metrics} \\
     Reward vs KL-Divergence & Obj. & \cite{cideron2024musicrl} \\
     Frechet Distance & Obj. & \cite{liao2024baton} \\
     KL-Divergence & Obj. & \cite{liao2024baton} \\
     Kolmogorov-Smirnov Test & Obj. & \cite{wu2023interval} \\

    \textbf{Content Accuracy} \\
     Word Error Rate & Obj. & \cite{anastassiou2024seed, chu2024qwen2, zhang2024speechalign, nagpal2025speech, huang2025step, luo2025openomni, gao2025emo} \\
     Speaker Similarity & Obj. & \cite{anastassiou2024seed, zhang2024speechalign, gao2025emo} \\
     LLM Evaluation & Obj. & \cite{chu2024qwen2, nagpal2025speech, huang2025step} \\
     Character Error Rate & Obj. & \cite{huang2025step} \\
     Short Time Objective Intelligibility & Obj. & \cite{kumar2025using} \\
         
    \textbf{Music} \\
     Harmonic Quality & Obj. & \cite{wu2025adaptive} \\
     Synchronization & Obj. & \cite{wu2025adaptive} \\
     Rhythmic Diversity & Obj. & \cite{wu2025adaptive} \\
              
     \textbf{Emotion} \\
     Emotion Change Consensus & Obj. & \cite{wu2023interval} \\
              
     \textbf{Generation} \\
     Cross-Modal Language-Audio Perceptual & Obj. & \cite{liao2024baton} \\
     Inception Score & Obj. & \cite{liao2024baton} \\

    \textbf{Human Evaluation} \\
     Mean Opinion Score & Subj. & \cite{yang2010ranking, liu2021reinforcement, cideron2024musicrl, anastassiou2024seed, wu2025adaptive, liao2024baton, kumar2025using, zhang2024speechalign, nagpal2025speech, huang2025step, gao2025emo} \\
     Human A/B Comparisons & Subj. & \cite{liu2021reinforcement, gao2025emo} \\
     HCI (Semi-Structured Interview/Creative Support Index) & Subj. & \cite{zhou2021interactive} \\
     \bottomrule
  \end{tabular}
\end{table}
\end{document}